\documentclass[]{spie}  

\usepackage{amsmath,amsfonts,amssymb}
\usepackage{graphicx}
\usepackage[colorlinks=true, allcolors=blue]{hyperref}
\usepackage{footnote} 
\makesavenoteenv{tabular}
\usepackage{wrapfig} 

\title{Development of the ROSIE Integral Field Unit on the Magellan IMACS Spectrograph}

\newcommand{\um}{$\mu$m}
\newcommand\arcsec{\mbox{$^{\prime\prime}$}}%
\newcommand\arcmin{\mbox{$^{\prime}$}}%
\newcommand{\degree}{\mbox{$^{\circ}$}}%

\author[a]{Rosalie C. McGurk}
\author[a]{Stephen A. Shectman}
\author[a]{Leon Aslan}
\author[b]{Chung-Pei Ma}
\affil[a]{Observatories of the Carnegie Institution for Science, 813 Santa Barbara Street, Pasadena, CA 91101, USA}
\affil[b]{Department of Astronomy, Department of Physics, University of California, Berkeley, CA 94720, USA}

\authorinfo{Further author information: (Send correspondence to R.C.M.)\\R.C.M.: E-mail: rmcgurk@carnegiescience.edu or rosalie.mcgurk@gmail.com, Telephone: 1 626 304 0203}

\begin{document} 
\maketitle

\begin{abstract}
We are building an image slicer integral field unit (IFU) to go on the IMACS wide-field imaging spectrograph on the Magellan Baade Telescope at Las Campanas Observatory, the Reformatting Optically-Sensitive IMACS Enhancement IFU, or ROSIE IFU. The 50.4\arcsec~$\times$~53.5\arcsec~field of view will be pre-sliced into four 12.6\arcsec~$\times$~53.5\arcsec~subfields, and then each subfield will be divided into 21 0.6\arcsec~x 53.5\arcsec~slices. The four main image slicers will produce four pseudo-slits spaced six arcminutes apart across the IMACS f/2 camera field of view, providing a wavelength coverage of 1800 Angstroms at a spectral resolution of 2000. Optics are in-hand, the first image slicer is being aluminized, mounts are being designed and fabricated, and software is being written. This IFU will enable the efficient mapping of extended objects such as nebulae, galaxies, or outflows, making it a powerful addition to IMACS.
\end{abstract}


\section{INTRODUCTION}
\label{sec:intro}  

Integral field spectroscopy, in which a spectrum is obtained at every position over a field of view, is a powerful tool for studying extended objects of all kinds, including but not limited to the stellar dynamics of galaxies, outflows and inflows of gas around galaxies and active black holes, resolved stellar populations and metallicity measurements, gravitationally lensed arcs, and pre-main sequence stellar objects. The advent of wide field integral field spectrographs with fields larger than 20\arcsec~x 20\arcsec on large telescopes, such as the Multi Unit Spectroscopic Explorer (MUSE) \cite{2010Bacon} on the Very Large Telescope or the Keck Cosmic Web Imager (KCWI) \cite{2018Morrissey}, has created the opportunity to more efficiently map faint and extended objects in one observation. 

Our image slicing integral field spectrograph design is based on the designs described by Content in Refs.~\citenum{1997Content, 1998Content}.  Instead of directing the light into clusters of fibers or lenslets, the image from the telescope is focused onto the image slicer mirror array, a stacked array of thin mirrors tilted at individual unique angles with respect to each other. The image slicer redirects slices of the image in multiple directions so that they can be reformatted into a new long slit and fed into a spectrograph to be dispersed. Software is then used to extract the spectra and reconstruct them into a 3-D datacube (x and y positions plus wavelength) of the original field.

Magellan currently lacks a wide-field integral field spectrograph similar to MUSE or KCWI. 
The need for this capability within the Magellan 6.5-m Telescope community has led to three integral field units (IFUs) or spectrographs currently under construction for Magellan: the IFU that we outline in this paper for the Inamori Magellan Areal Camera and Spectrograph (IMACS) \cite{2006Dressler,2011Dressler}, the Large Lenslet Array Magellan Spectrograph (LLAMAS, PI Rob Simcoe), and the Integral Field Unit for the Magellan/Michigan Fiber System (M2FS \cite{2012Mateo}) (IFU-M, PI Mario Mateo) .

Mounted on the Magellan Baade Telescope, IMACS is a wide field spectrograph delivering a 27.5\arcmin~field to its f/2.3 camera. IMACS has a large mask server located in the telescope focal plane; this mask server can hold up to 6 multi-slit masks, but also creates a space for other fixtures to modify the focal plane into different formats, such as the Gladders Image-Slicing Multislit Option (GISMO) reformatting unit for dense-pack spectroscopy \footnote{http://www.lco.cl/telescopes-information/magellan/instruments/imacs/gismo/gismoquickmanual.pdf}.  Placing such reformatting units in the mask server saves the cost of building an entire spectrograph from scratch and enables new science for an existing, permanently mounted instrument.   

The \textbf{R}eformatting \textbf{O}ptically-\textbf{S}ensitive \textbf{I}MACS \textbf{E}nhancing Integral Field Unit, or ROSIE IFU, will fit into the IMACS mask server.  When used with the IMACS f/2.3 camera with 0.2\arcsec~pixels, ROSIE IFU will provide a 50.4\arcsec~$\times$~53.5\arcsec~field of view that is sliced into 84 slices that are 0.6\arcsec~wide and 53.5\arcsec~long. Briefly, the optics in ROSIE split the field into four subfields, magnify the subfields by a factor of three, slice and fan the image of each subfield into 21 slices, magnify each slice by a factor of one third to return the pixel scale to its original size, and align each of the 21 slices into staggered pseudo slits that are sent through the IMACS spectrograph and dispersed onto the detector. The highest dispersion grisms for the f/2.3 camera will provide wavelength ranges of 3900-5650 \AA, 5300-8000 \AA, or 6720-9000 \AA. IMACS also has an f/4.5 camera with 0.11\arcsec~per pixel and higher spectral resolution gratings that may be used with ROSIE, but the field of view will be more than 60\%~smaller and custom wavelength band blocking filters will need to be purchased; this mode is currently unsupported but may be developed in the future.  For convenience, ROSIE IFU's specifications are also summarized in Table \ref{tab:specs}.

Section \ref{sec:optlayout}~discusses the optical layout of ROSIE IFU and the status of procuring its optics. Section \ref{sec:mechlayout}~discusses the development of the mechanical layout as well as current alignment plans. Section \ref{sec:imgslicer}~explains the novel development of our image slicers in depth. 

\begin{table}[t]
\caption{Specifications of the ROSIE IFU}
\label{tab:specs}
\begin{center}
\begin{tabular}{|l|l|} 
\hline
\rule[-1ex]{0pt}{3.5ex}  Field of View 		& 50.4\arcsec~x 53.5\arcsec  \\ \hline
\rule[-1ex]{0pt}{3.5ex}  Slice Width 		& 0.6\arcsec  \\ \hline
\rule[-1ex]{0pt}{3.5ex}  Slice Length 		& 53.5\arcsec  \\ \hline
\rule[-1ex]{0pt}{3.5ex}  Number of Slices 	& 84  \\ \hline
\rule[-1ex]{0pt}{3.5ex}  Maximum Spectral Resolution & 1800  \\ \hline
\rule[-1ex]{0pt}{3.5ex}  Highest Resolution Grisms & Wavelength Ranges Available  \\ 
\rule[-1ex]{0pt}{3.5ex}  400 lines/mm blaze 21.2\degree	& 3900-5650 \AA  \\ 
\rule[-1ex]{0pt}{3.5ex}  300 lines/mm blaze 17.5\degree	& 5300-8000 \AA  \\ 
\rule[-1ex]{0pt}{3.5ex}  300 lines/mm blaze 26.7\degree	& 6720-9000 \AA  \\ \hline
\end{tabular}
\end{center}
\end{table} 

\section{Optical Layout}
\label{sec:optlayout}

   \begin{figure} [ht]
   \begin{center}
   \begin{tabular}{c} \rule[-1ex]{0pt}{0.5ex} 
   \includegraphics[width=0.9\columnwidth]{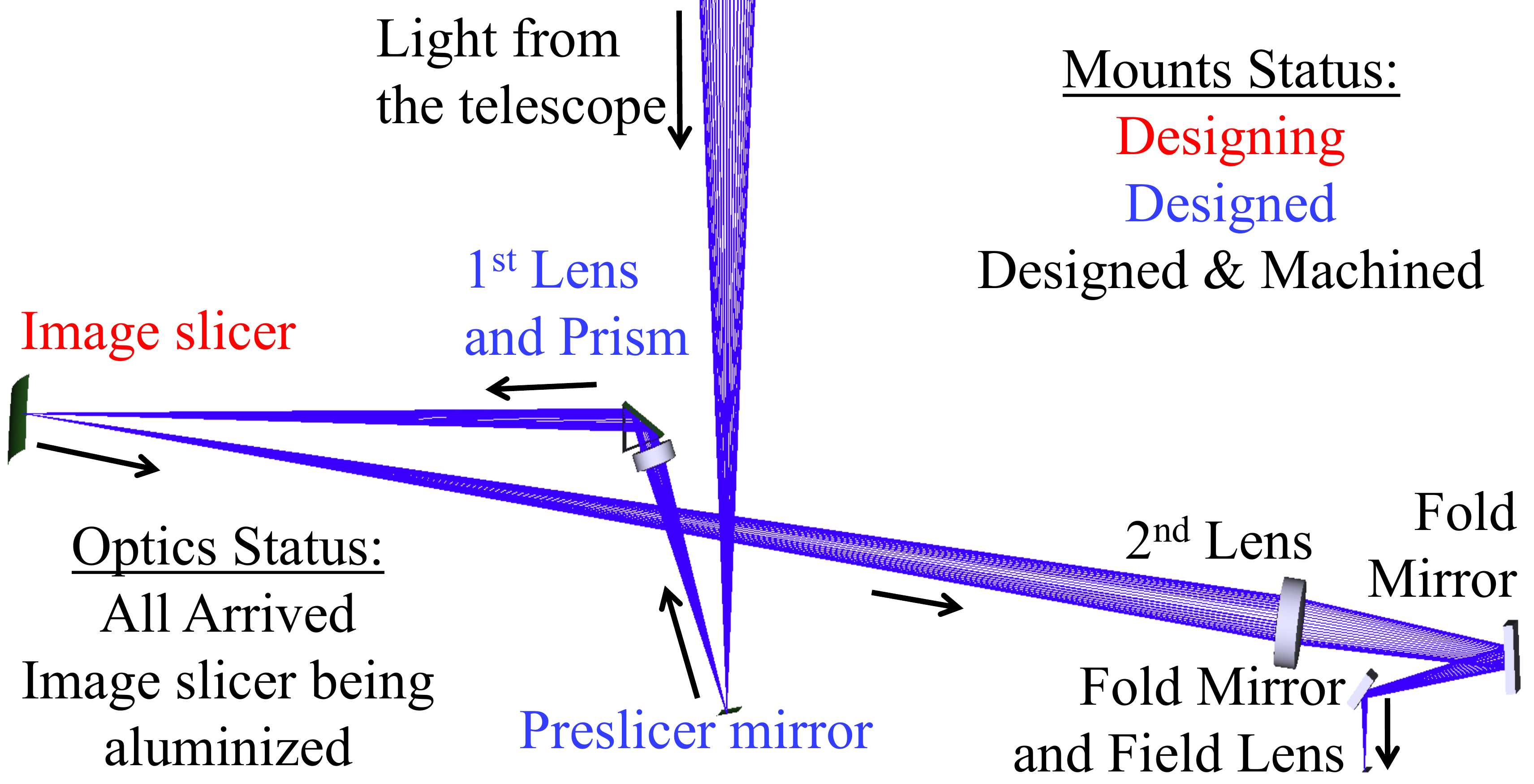} \\ \rule[-1ex]{0pt}{0.5ex} 
   \includegraphics[width=0.9\columnwidth]{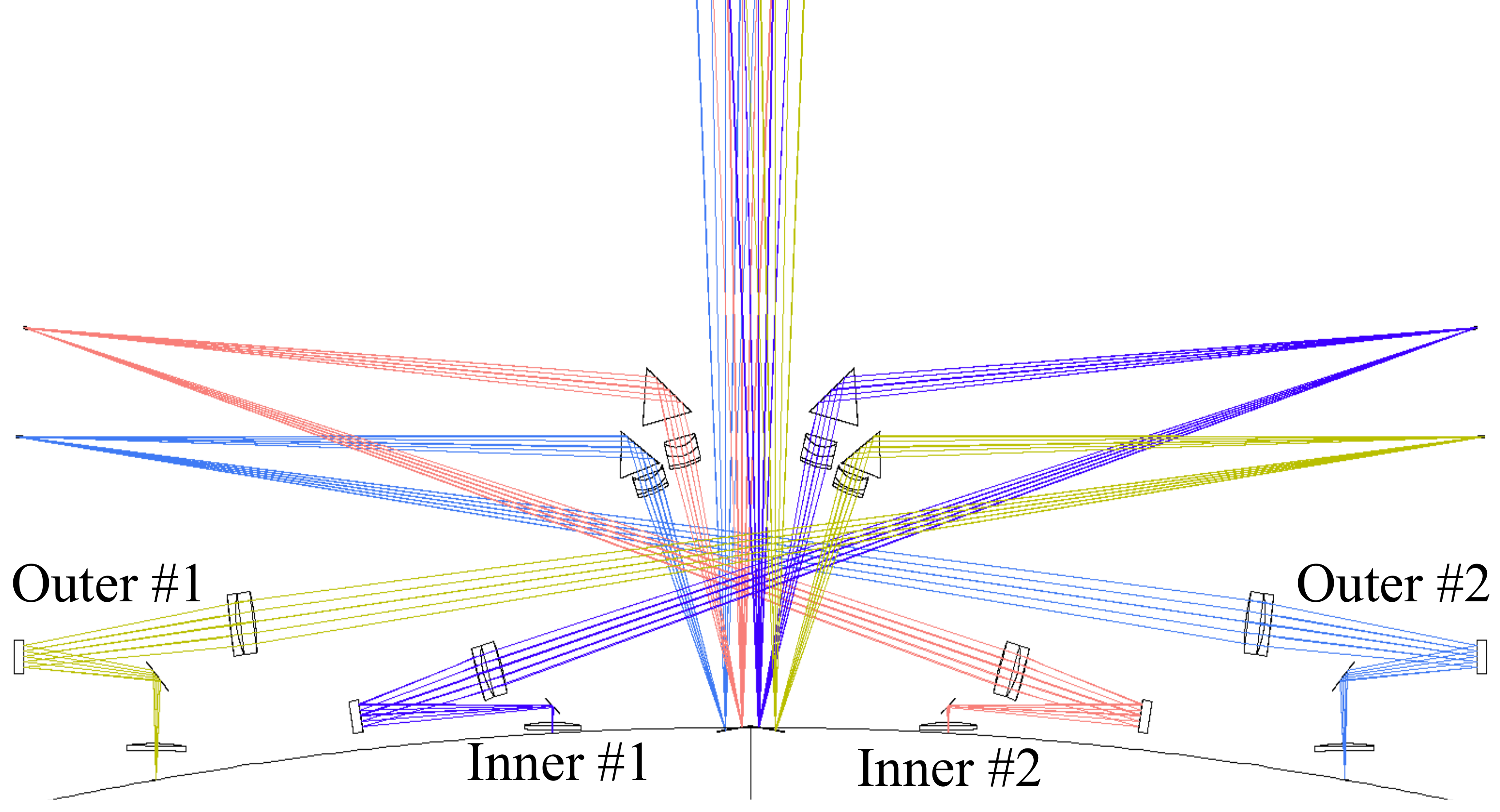} 
   \end{tabular}
   \end{center}
   \caption[opticallayout]    { \label{fig:opticallayout} 
\textbf{(top)}~Optical layout of the central slice of one outer subfield. The image slicer fans out the slices into and out of the page; we use the center slice to demonstrate the rest of the light path and optics that are present for each fanned beam. 
\textbf{(bottom)}~Optical layout of the central slice of four subfields. The field is pre-sliced into 4 subfields, and each subfield is sliced into 21 slices, fanned into and out of the page, making a total of 84 optical paths after slicing. Each subfield's 21 slices are reformatted into a pseudo slit to send into IMACS. }
   \end{figure} 

   \begin{figure} [ht]
   \begin{center}
   \begin{tabular}{c} \rule[-1ex]{0pt}{0.5ex} 
   \includegraphics[width=0.9\columnwidth]{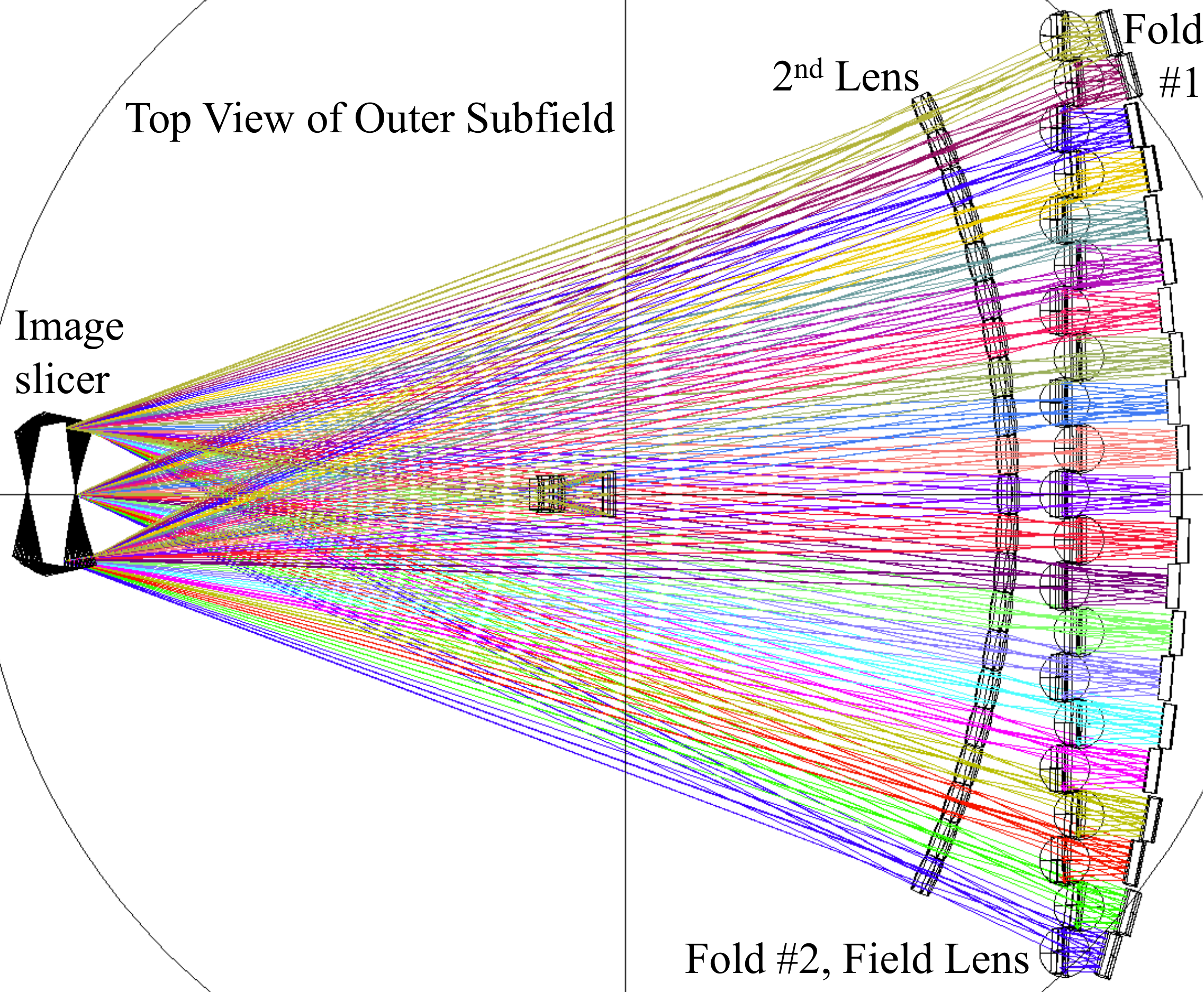} 
   \end{tabular}
   \end{center}
   \caption[topview]    { \label{fig:topview} 
Top view of the optical layout of one outer subfield; this is a top view, or view from the light incoming from the telescope, of the optical layout shown in the top panel in Figure \ref{fig:opticallayout}. The image slicer slices the subfield into 21 beams, each shown in a different color, which fan out in 1.19\degree~increments. The fanned beams are each magnified by a factor of 1/3 by the second achromatic doublet lens, and then fold mirrors \#1 and \#2 reposition the light to go down through the field lens into the spectrograph into the page.  }
   \end{figure}

ROSIE IFU's optical design is particularly challenging because of the tight space constraints dictated by needing to fit the IFU inside the IMACS mask server. ROSIE IFU's cylindrical volume can be no larger than 546 mm circular diameter by 240 mm height; since ROSIE uses a total of 342 traditional optics (generally smaller than 1 inch) plus four image slicers with 21 slices to slice and reformat the field of view, designing the optics and mounts required close attention to detail.  The two inner subfields are identical in design to each other; the second inner subfield is located and rotated 180\degree~around the center z-axis of the telescope focal plane from the first. Similarly, the two outer subfields are identical to one another and rotationally symmetrical around the center of the focal plane. 

Figure \ref{fig:opticallayout}~illustrates the layout for the center slice of one outer subfield, and for all four subfields. A four-segment preslicer is placed at the telescope focal plane. The four facets of the preslicer are cut from a single concave mirror with a radius of curvature of 220.0 mm. The facets are approximately 19.2 mm $\times$~5.1 mm in size, and in combination with the first achromat form separate pupil images 6.0 mm in diameter at a distance of 110.0 mm.  Near each pupil image is a small cemented optical doublet made of S-NSL36 and S-FPL51 glass which re-images the telescope focal plane from the preslicer facet onto the main image slicer at a magnification of 3:1, and a total internal reflection fold prism made of fused silica which deflects the light through an angle of 104.4\degree~(inner subfields) or 110\degree~(outer subfields) toward one of the four main slicers. Use of the total internal reflection fold prism avoids possible polarization of the light that could have occurred if a normal flat fold mirror was used with such a large angle of incidence.

\begin{figure} [ht]
   \begin{center}
   \begin{tabular}{c}  
   \includegraphics[width=0.9\columnwidth]{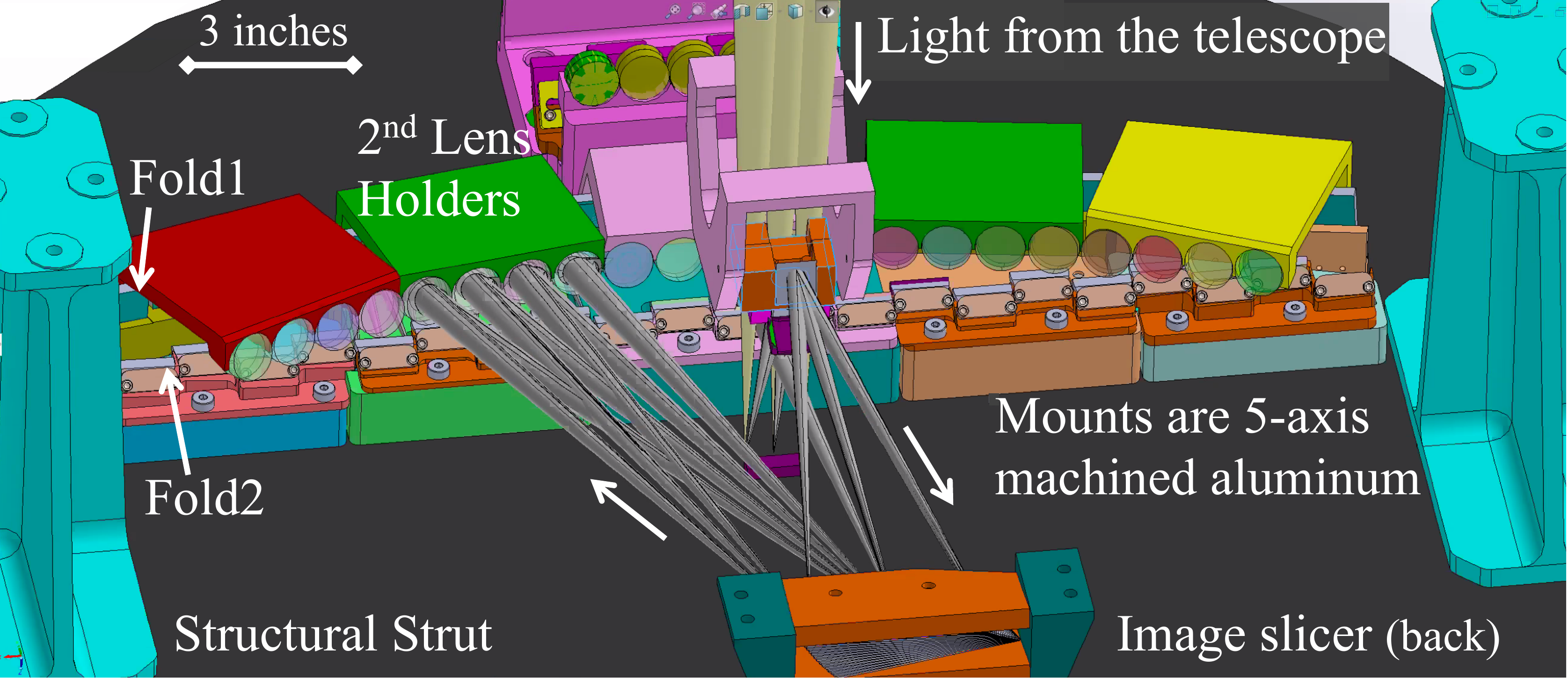} 
   \end{tabular}
   \end{center}
   \caption[mechanicallayout]    { \label{fig:mechanicallayout} 
Mechanical layout of the mounts designed for both inner subfields, although only one inner subfield is shown for clarity.}
   \end{figure}

Our four image slicers are each made of 21 fused quartz slices that are 618 \um~thick, 64.0 mm long, and 28.0 mm wide. The active optical surface of each slice is on the 0.618 mm $\times$~64.0 mm edge and is figured with a concave radius of curvature, 250.0 mm for the inner image slicers and 251.7 mm for the outer image slicers, to form pupil images 8.1 mm and 10.5 mm, respectively, in diameter at distances of 384 mm and 418 mm from the image slicers.  The slices for the inner and outer subfields' image slicers are tilted with respect to their neighboring slices in increments of 1.52\degree~and 1.19\degree, respectively.  If the optical layout of Figure \ref{fig:opticallayout}~is viewed from the top, we can see the 21 outer subfield beams fanning out from the image slicer in Figure \ref{fig:topview}.  We show four beams fanning out from the image slicer to the mounted second achromatic doublet lenses in Figure \ref{fig:mechanicallayout}. The construction of the image slicers is described in detail in Section \ref{sec:imgslicer}. 

\begin{figure} [ht]
   \begin{center}
   \begin{tabular}{c} \hspace{-1em} 
   \includegraphics[width=.99\columnwidth]{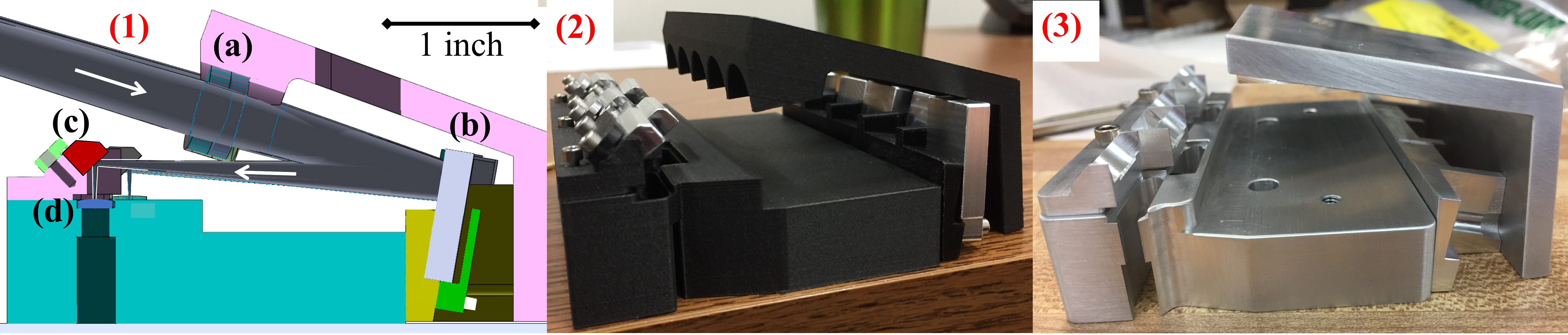}  
   \end{tabular}
   \end{center}
   \caption[centermount]    { \label{fig:centermount} 
\textbf{(1)}~Model of the mounting fixture for the central five opticalbeams. Each fanned beam requires the various optics to be held at unique angles from one another. The following optics are labeled: (a) the second achromatic doublet lens, (b) fold mirror \#1, (c) fold mirror \#2, and (d) field lens.
\textbf{(2)}~3-D printed plastic mount model to test alignment procedures. These tests confirmed that 3-D printed mounts can not achieve the accuracy we needed. 
\textbf{(3)}~5-axis machined aluminum mount. Optical alignment is underway.
}
   \end{figure}

Near the pupil image of each slice is a cemented achromatic doublet made of S-FPL51 and S-NSL36 glass which re-images the image slicer onto the telescope's curved focal plane (convex with a curvature of 1233.2 mm) at a magnification of 1:3, to return the beams to their original focal ratio, f/11.  Located after the achromat, a pair of fold mirrors reposition the focused slices of each subfield into a pseudo-slit 400 mm long, with heights that vary along the telescope's curved focal plane. While the first fold mirrors have no space constraints, the second fold mirrors are located very near the fanned beams overhead and very close to their neighboring mirrors, so we designed them to have custom half-circle shapes with shaved top edges, as illustrated in Figure \ref{fig:centermount}. Since four parallel 400 mm long slits would lower the structural integrity of the baseplate upon which most of the optics are mounted, the neighboring slices are offset from each other by alternating +/- 3mm steps from each pseudo slit's center.  Finally, each beam has a 2 mm thick field lens located 1 mm and 8 mm, respectively, before the focus of each slice in the inner and outer subfields to restore the exit pupil of the reformatted slit image to the distance of the image of the telescope primary mirror formed by the telescope secondary. Due to the varying angles of incidence on slices of the image slicers and how it changes the pupil locations in the more widely fanned beams, the field lenses in the inner subfields need three different convex radii of curvature and the field lenses in the outer subfields need another three different radii of curvature. The optical prescription for the inner and outer subfields is given in Table \ref{tab:optics}.

The optical design of the entire system is complete. We are building the 1st inner subfield as a proof of concept, and then will proceed to build the other three subfields.  We have all of the traditional optics (achromatic doublet lenses, flat and curved fold mirrors, and triangular fold prism) in hand. The first image slicer is nearly completed; the holes have been drilled in the slices, and the slices have been polished with the concave 250.0 mm spherical surface. The slices are being cleaned and placed into the aluminizing assembly, and will be sent to be aluminized in the next week. See Section \ref{sec:imgslicer}~for more details about constructing the image slicers.

\section{Mechanical Layout}
\label{sec:mechlayout}

As shown in Figure \ref{fig:mechanicallayout}, the mounts and mechanical layout is mostly complete for the optics of the two inner subfields.  The mounts for the system can be divided into two categories: adjustable and fixed. The triangular fold prism and main image slicers need to be adjustable along two axes to allow slight tip-tilt adjustments to the outgoing beams; the first achromatic doublet has a small range of tilt adjustment about its central long axis. 

The preslicer and all of the optics after the image slicer (second achromatic doublet, fold mirrors, and field lens) are fixed into position.  Each fanned beam requires the various optics to be held at unique angles from one another, and our limited space requirements dictate that multiple beam configurations must be mounted together in units of four to five beams. To achieve the multiple unique angles required, we purchased a 5-axis tilt-rotary add-on to our 3-axis computer numerical control (CNC) milling machine in order to construct these parts in-house. The 5-axis machining in aluminum provides precise location and orientation of reference mating surfaces for the optical components. We are prototyping the mount of the central five fanned beams to test our alignment plan. The layout of this mount is shown in panel (1) of Figure \ref{fig:centermount}. The second achromatic doublets and field lenses will be bonded into their mounting pockets with room temperature vulcanizing (RTV) silicone. Both fold mirrors will  be held in place against 3 mounting pads by a wave spring and a clamp; we will also apply a few beads of RTV silicone after alignment is complete to secure the optics against any physical shocks.  While waiting for the 5-axis tool to arrive, we 3-D printed the mount out of plastic, as shown in Figure \ref{fig:centermount}~panel (2). This system has been useful to practice installing the optics and to test alignment procedures, but our 3-D printer cannot create sufficiently accurate surfaces for the optic mounts. Since then, the 5-axis tool has arrived, and we now have machined aluminum mounts in which to begin mounting optics.

\section{Image Slicer Construction}
\label{sec:imgslicer}

One of the most difficult and critical parts of an image slicing IFU is the image slicer proper. Each slice must be figured to the appropriate surface shape and assembled at a precise position and angle with respect to its neighbors.

   \begin{figure} [t]
   \begin{center}
   \begin{tabular}{c} \hspace{-1em} 
   \includegraphics[width=.99\columnwidth]{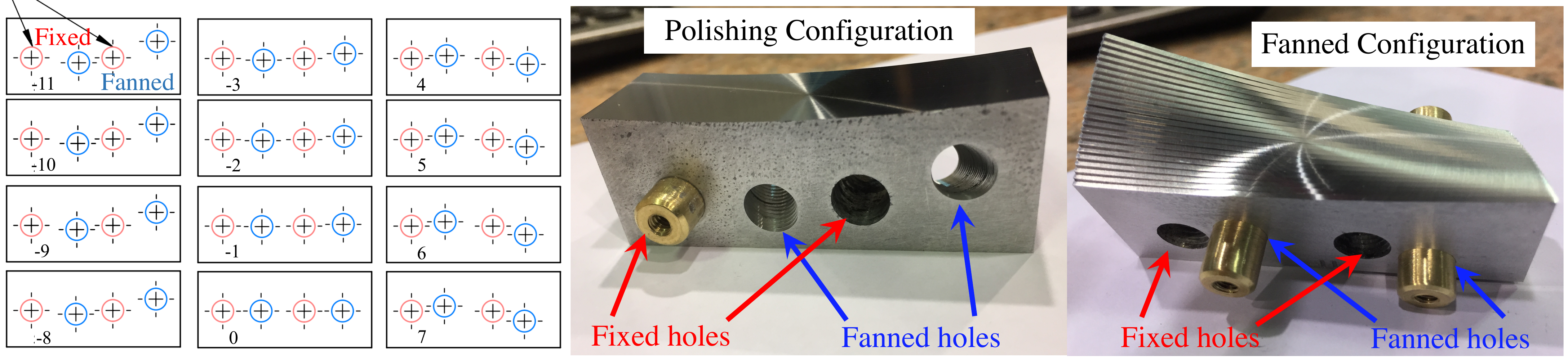}  
   \end{tabular}
   \end{center}
   \caption[fixedfanned]    { \label{fig:fixedfanned} 
\textbf{(left)} Drawings of 12 slices (out of 21) showing the four hole positions. The two red holes are fixed, or have the same position on each slice. The two varying blue holes are fanned, or have positions that rotate around the central slice's vertex. The slices are numbered with their position with respect to the center slice (\#0). 
\textbf{(center)} Polishing assembly as demonstrated by our aluminum slice prototypes. The fixed/fanned holes are indicated with red/blue arrows, respectively. By placing pins through the fixed holes, the slices form into a rectangular block, ready to be polished. 
\textbf{(right)} Fanned assembly as demonstrated by our aluminum slice prototypes. By placing pins through the fanned holes, the slices are held at precise tilt angles with respect to their neighbors.
} 
   \end{figure} 

   \begin{figure} [ht]
   \begin{center}
   \begin{tabular}{c} \hspace{-1em}  
   \includegraphics[width=.99\columnwidth]{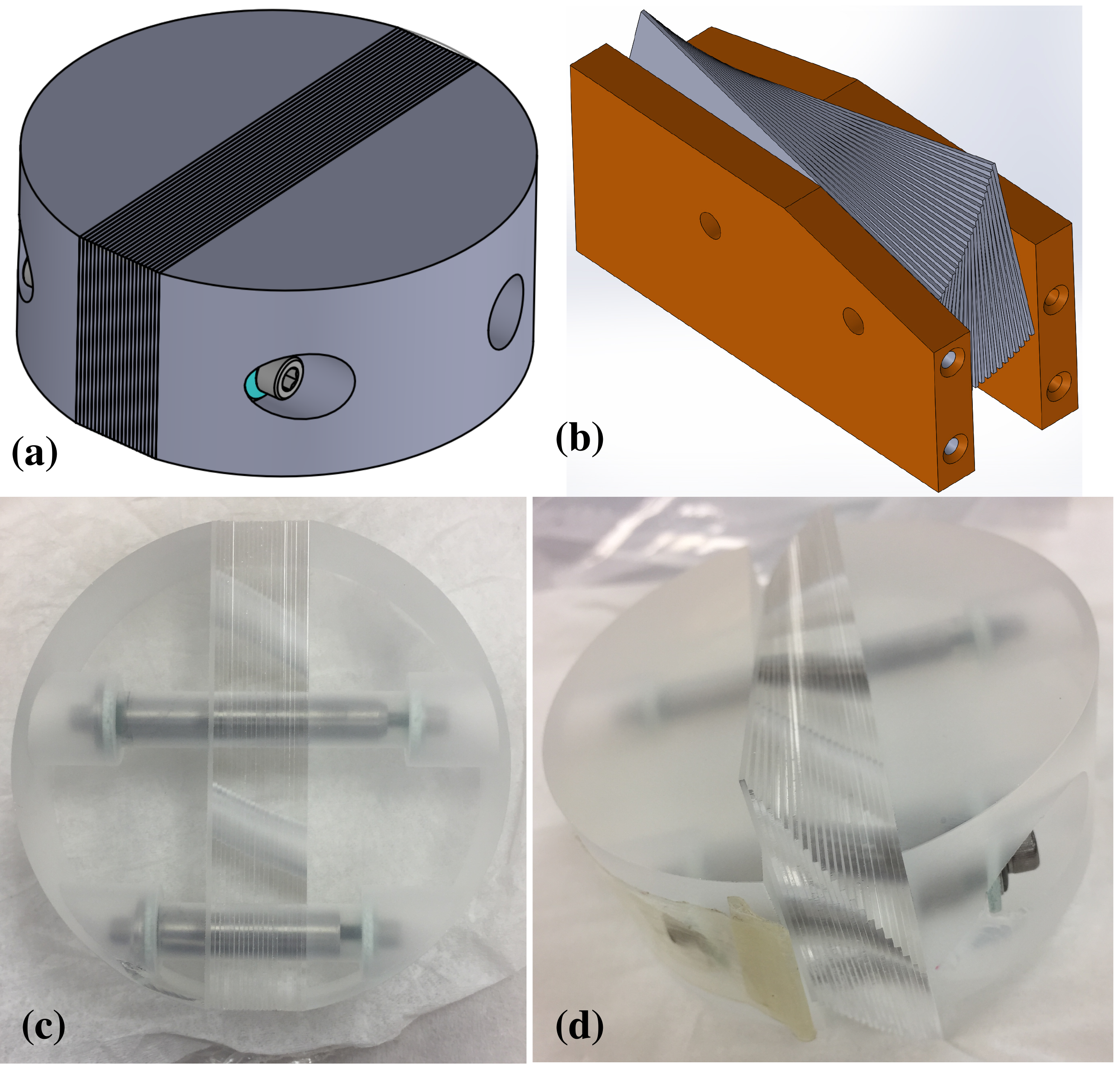}  
   \end{tabular}
   \end{center}
   \caption[modelquartz]    { \label{fig:modelquartz} 
\textbf{(a)} Model of the image slices and round slice holders ready to polish a concave 250 mm radius of curvature onto the front surface. As a reference, the 21 slices stacked are 12.98 mm or 0.52 inches thick, smaller than a dime's diameter of 17.9 mm or 0.71 inches. 
\textbf{(b)} Model of the fanned image slicer with a simple assembly mount.
\textbf{(c)} Image of the image slicer and slice holders assembled in wax and ready for polishing. 
\textbf{(d)} Before assembling the slices into the polishing configuration, we assembled the slices into the fanned configuration to confirm the positioning of the fanned holes. Note the even spacing of the slice edges in the image foreground. 
} 
   \end{figure}

We are using an innovative and cost-saving method to make the image slicers. The prevalence of fused quartz or silicon wafers in the semiconductor industry means that custom thickness wafers with 5 \um~thickness tolerance are easily available and affordable from which to make the slices. Each slice has four precise holes drilled into its flat surface, as shown in Figure \ref{fig:fixedfanned}. Two of the holes, called fixed holes, are in the same place on each slice; we use centerless ground invar pins through the fixed holes to hold the slices together as a block or slab with no tilts between slices. Once round slice holders are added to the block of slices, as shown in the central panel of Figure \ref{fig:fixedfanned}~and panels (a) and (c) of Figure \ref{fig:modelquartz}, it resembles a cylinder or puck and is relatively easy to polish with a spherical surface and, with extra spacing between each slice, to aluminize.  The second two holes, called fanned holes, are different on each slice; their positions rotate with different radii around the vertex of the central slice. By placing invar pins through the fanned holes, the slices are held at precise tilts with respect to their neighbors, as shown in the right panel of Figure \ref{fig:fixedfanned}~and panels (b) and (d) of Figure \ref{fig:modelquartz}. The slices for the inner and outer subfield image slicers fan in increments of 1.52\degree~and 1.19\degree, respectively.

\begin{figure} [ht]
   \begin{center}
   \begin{tabular}{c} \hspace{-1em}  
   \includegraphics[width=.99\columnwidth]{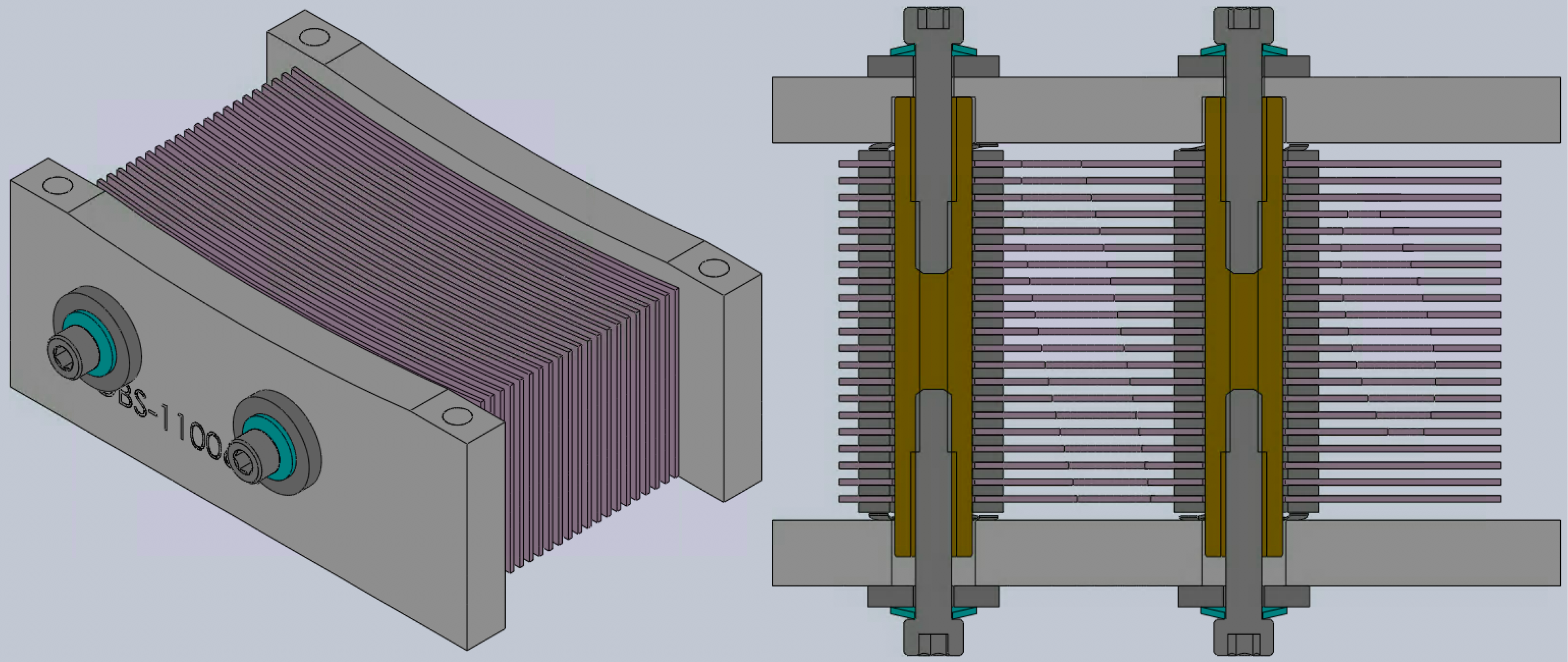}  
   \end{tabular}
   \end{center}
   \caption[alumassem]    { \label{fig:alumassem} 
\textbf{(left)} Model of the aluminizing assembly. The slices, shown in light purple, are spaced along invar pins by 1mm thick ring shims, and bracketed by light gray invar plates. 
\textbf{(right)} Cross section of the aluminizing assembly, revealing the invar pins, ring shims, wave washers, washers, and vented screws sandwiching the slices between the invar plates.} 
   \end{figure}

The construction process is summarized here, with details to follow:
\begin{enumerate}
\item Acquire wafers of the correct thickness, cut slices to size, and drill holes.
\item Assemble the slices in a hot wax bath using pins through the fixed holes with round holders on either end to create a circular optic to be polished (polishing assembly as seen in panels (a) and (c) of Figure \ref{fig:modelquartz}).
\item Polish to correct spherical shape.
\item Take the polishing assembly apart in hot water bath to melt the wax and clean the slices.
\item Assemble the slices using pins through the fixed holes, spaced apart with ring shims, for aluminizing (aluminizing assembly as seen in Figure \ref{fig:alumassem}).
\item Aluminize.
\item Assemble the slices using precise diameter invar pins through the fanned holes (un-aluminized fanned assembly shown in panels (b) and (d) of Figure \ref{fig:modelquartz}).
\end{enumerate}

   \begin{figure} [ht]
   \begin{center}
   \begin{tabular}{c} \hspace{-1em} 
   \includegraphics[width=.99\columnwidth]{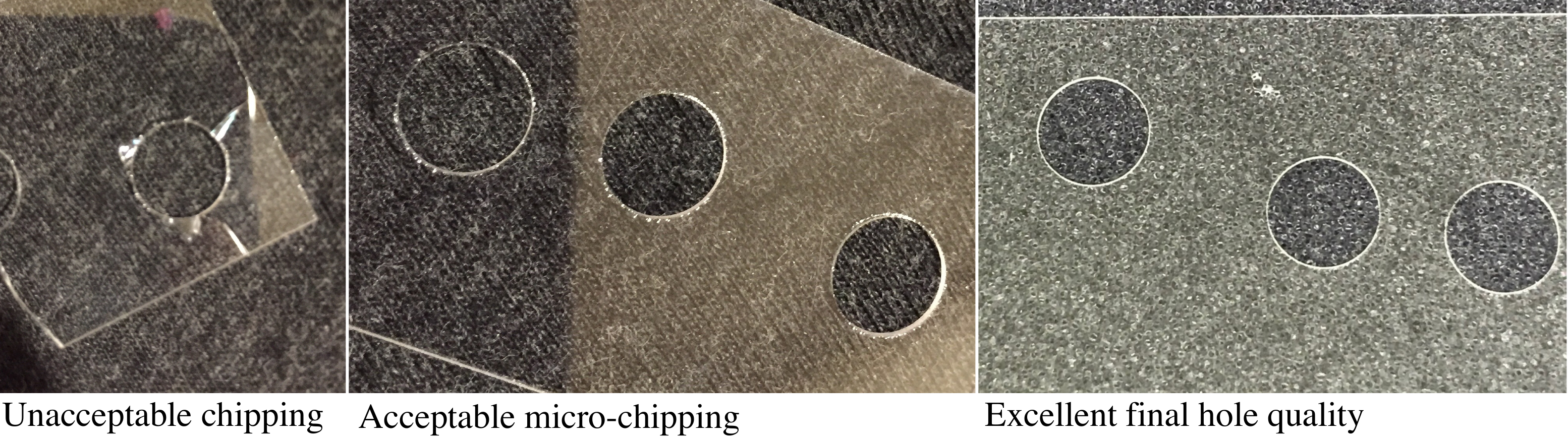}  
   \end{tabular}
   \end{center}
   \caption[holequality]    { \label{fig:holequality} 
Drilling holes with little-to-no chipping in 618\um~thick fused quartz wafers is possible once wafers are stacked. \textbf{(left)} If wafers are drilled one at a time, high levels of chipping and breakage can result. \textbf{(middle)} and \textbf{(right)}: Once wafers are stacked, holes have either small amounts of microchipping or almost none.} 
   \end{figure} 

   \begin{figure} [ht]
   \begin{center}
   \begin{tabular}{c} \hspace{-1em} 
   \includegraphics[width=.99\columnwidth]{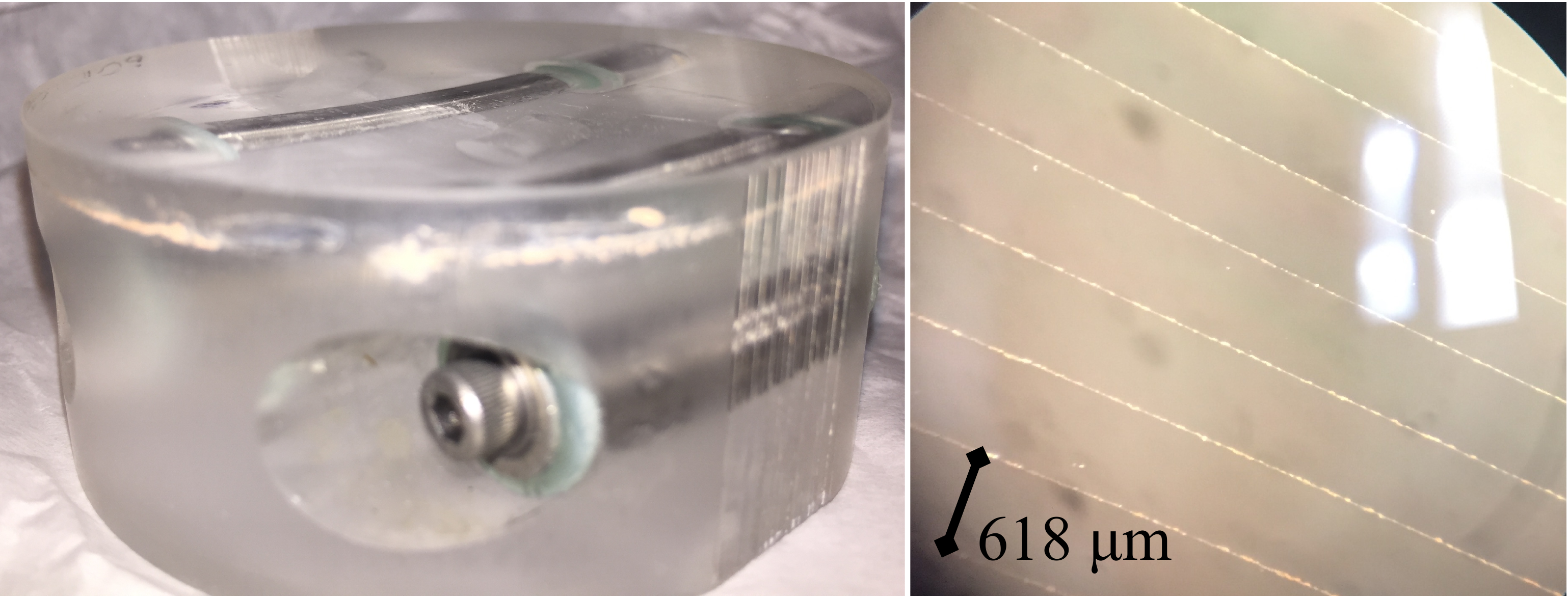}  
   \end{tabular}
   \end{center}
   \caption[polishzoom]    { \label{fig:polishzoom} 
\textbf{(left)} Image of the polished image slicer and holders. The curvature of the top surface is visible in the apparent bending of the straight invar pin.
\textbf{(right)} Microscopic view of the polished slices (each 618 \um~thick). There is very little chipping at the slice boundaries. Dark specks are dust on the microscope lenses.} 
   \end{figure}

Each of our image slicers is made of 21 fused quartz slices that are 618 \um~thick, 64.0 mm long, and 28.0 mm wide.  Insaco Inc.\footnote{\url{https://www.insaco.com/}}, a company that specializes in high precision machining of hard and brittle materials such as quartz, glass, and ceramics, drilled the holes with 5 \um~diameter tolerance and 10 \um~true position tolerance. After some experimentation, Insaco found that drilling stacks of slices drastically decreased the amount of chipping around the holes and eliminated breakage of all stacked slices except the bottom one. The left panel of Figure \ref{fig:holequality}~shows the chipping and breakage from drilling a single slice at a time, while the middle and right panels clearly illustrate the low amounts of microchipping around the holes once slices were stacked during drilling.  Insaco also fabricated the round slice holders shown in panels (c) and (d) of Figure \ref{fig:modelquartz}~out of fused quartz, so that the entire polishing assembly would be made from the same material and polish uniformly.

To assemble the slices into the polishing configuration, we put invar pins through the first round slice holder, the fixed holes of the slices in the correct order, and the second slice holder, and secured the assembly together with vented screws through a large fiberglass washer, an aluminum washer, and a Belleville washer. The Belleville washers allow us to control the clamping force of the assembly in order to avoid breakage. We chose to complete the entire assembly process in 60-75\degree~C paraffin wax in order to seal the spaces between the slices against contamination during polishing. We also hoped that sealing the slice boundaries would decrease the amount of chipping along the slice edges. When we inspect the slice boundaries by microscope after polishing, shown in the right panel of Figure \ref{fig:polishzoom}, the edges are in excellent condition. Based on width measurements before and after assembly with wax, the wax layer between each slice is 1-2 \um~thick, and therefore does not offset the slices enough with their surface curvature of 250 mm to create much of a wedge difference in the slice surface. An unforeseen but surmountable challenge is that the slices and slice holders have the same index of refraction as the melted paraffin wax, making them nearly invisible when immersed in the wax bath; we recommend Sharpie marker pen dots along the slice edges to make the slices visible. Polishing was done by Kreischer Optics\footnote{\url{https://www.kreischeroptics.com/}}, and looks excellent as seen in Figure \ref{fig:polishzoom}.

After it is polished, warming the polishing assembly in hot water and disassembling it was straightforward. We designed the pins to slide out of the slice holders and slices; we removed one pin fully and then slowly retracted the second pin from the assembly as each subsequent slice was removed, in order to keep the slices in their proper order. The melted paraffin wax acted as a lubricant and made it easy to slide each slice off its neighbor. We gently cleaned residual wax off the slices while they were still warm, but a layer of wax remained. A 50:50 mixture of kerosene and acetone is extremely effective at removing the paraffin wax, but is also effective at eating through nitrile and latex gloves; we recommend chemical-resistant tightly-fitted neoprene gloves. Our kerosene-acetone cleaning procedure involved soaking each slice in the kerosene-acetone mixture for several minutes, wiping the slice surfaces with gauze while held flat, cleaning the holes with a cotton swab, wiping the slice edges with a gauze pad, rinsing the kerosene-acetone mixture off with soapy water and gentle wipes, rinsing the soapy water off with distilled water, and drying. We are also wiping the slices down with acetone and then distilled water to clean the slices one final time before assembling them to be aluminized.
  
Each slice needs to be aluminized independently from its neighbors; if the slices were coated in a block or in the fanned configuration, respectively, the coating would tear when the slices were separated or would be uneven due to shadowing by their neighboring slices. We have designed an aluminizing assembly as shown in Figure \ref{fig:alumassem}, held together by invar pins and invar plates. The slices are separated by 1.0 mm thick stainless steel ring shims with 14.0 mm outer diameters that are well clear of the slice surfaces, and the stack of shims plus slices is bracketed by wave washers on both ends of the invar pins. The use of the wave washer plus the somewhat undersize invar pin diameters allow for possible expansion during the coating process without threatening to stress or crack the fused quartz slices. Evaporated Coatings Inc.\footnote{\url{https://www.evaporatedcoatings.com/}} is coating our slices with a protected aluminum coating. Once our slices return coated, we will be able to disassemble the aluminizing assembly and assemble the completed image slicer using invar pins with precise diameters placed through the fanned holes.

\section{Final Thoughts}

Optical alignment and mounting is proceeding, as is the machining of additional mounts and the assembly of the image slicer to be aluminized. 


\acknowledgments 
 
The authors acknowledge the generous support of Carnegie Observatories, Las Campanas Observatory, and the Heising-Simons Foundation (2018-0768, C.-P. M.). 
The ROSIE IFU team also thanks Carnegie Observatories Director John Mulchaey, the Carnegie Instrumentation Team, and the Carnegie Machine Shop staff.

\begin{table}[t]
\caption{\textbf{Optical Design of the ROSIE IFU}}
\label{tab:optics}
\begin{center} \vspace{-7.72pt}
\begin{tabular}{|l|c|c|} 
\hline
\rule[-1ex]{0pt}{2.5ex} \textbf{Parameter}  									& \textbf{Inner subfields} 	& \textbf{Outer subfields}   \\ \hline
\rule[-1ex]{0pt}{2.5ex} Preslicer Radius of Curvature						& \multicolumn{2}{|c|}{concave 220.0 mm}  \\ \hline
\rule[-1ex]{0pt}{2.5ex} Preslicer Surface Tilt Angle							& 7.0\degree 	& 9.56\degree \\ \hline
\rule[-1ex]{0pt}{2.5ex} Preslicer Physical Size								& \multicolumn{2}{|c|}{19.2 mm $\times$~5.1 mm}  \\ \hline
\rule[-1ex]{0pt}{2.5ex} Distance betw.~Preslicer and First Achromatic Doublet\hspace{-1em} 		&  73.84 mm		& 68.91 mm  \\ \hline
\rule[-1ex]{0pt}{2.5ex} First Achromatic Doublet Radius of Curvature 1  		&  convex 25.91 mm	& convex 24.43 mm  \\ \hline
\rule[-1ex]{0pt}{2.5ex} First Achromatic Doublet Radius of Curvature 2  		&  convex 9.85 mm		& convex 9.53 mm  \\ \hline
\rule[-1ex]{0pt}{2.5ex} First Achromatic Doublet Radius of Curvature 3		&  flat			& flat  \\ \hline
\rule[-1ex]{0pt}{2.5ex} First Achromatic Doublet First Lens Thickness		&  3.0 mm		& 1.5 mm  \\ \hline
\rule[-1ex]{0pt}{2.5ex} First Achromatic Doublet Second Lens Thickness	&  5.0 mm 		& 4.0 mm  \\ \hline
\rule[-1ex]{0pt}{2.5ex} First Achromatic Doublet First Lens Material 			&  \multicolumn{2}{|c|}{S-NSL36}   \\ \hline
\rule[-1ex]{0pt}{2.5ex} First Achromatic Doublet Second Lens Material 		&  \multicolumn{2}{|c|}{S-FPL51}   \\ \hline
\rule[-1ex]{0pt}{2.5ex} First Achromatic Doublet Diameter					&  15.0 mm 	& 16.1 mm  \\ \hline
\rule[-1ex]{0pt}{2.5ex} Distance betw.~First Achromatic Doublet and Prism 	& 6.08 mm  	& 1.82 mm  \\ 
\rule[-1ex]{0pt}{2.5ex} Isosceles Angles of Triangular Fold Prism			&  54.4\degree	& 55.0\degree  \\ \hline
\rule[-1ex]{0pt}{2.5ex} Base Width of Triangular Fold Prism					&  18.45 mm 	& 14.66 mm  \\ \hline
\rule[-1ex]{0pt}{2.5ex} Isosceles Face Width of Triangular Fold Prism		&  15.85 mm 	& 12.78 mm \\ \hline
\rule[-1ex]{0pt}{2.5ex} Length of Triangular Fold Prism						&  13.00 mm 	& 13.00 mm  \\ \hline
\rule[-1ex]{0pt}{2.5ex} Distance betw.~Triangular Fold Prism and Image Slicer\hspace{-1em} 		& 197.00 mm		& 191.00 mm  \\ \hline
\rule[-1ex]{0pt}{2.5ex} Image Slicer Radius of Curvature					&  \hspace{-1em}concave 250.00 mm\hspace{-1em}	& \hspace{-1em}concave 251.70 mm\hspace{-1em}  \\ \hline
\rule[-1ex]{0pt}{2.5ex} Vertical Tilt Angle of Image Slicer						&  6.0\degree 			& 3.3\degree  \\ \hline
\rule[-1ex]{0pt}{2.5ex} Horizontal Tilt Angle Increments between Slices		&  1.52\degree 			& 1.19\degree  \\ \hline
\rule[-1ex]{0pt}{2.5ex} Distance betw.~Image Slicer and Second Achromatic 		& 321.73 mm	& 391.49 mm \vspace{-1pt} \\ 
\rule[-1ex]{0pt}{2.5ex} ~ ~Doublet		&  	&   \\ \hline
\rule[-1ex]{0pt}{2.5ex} Second Achromatic Doublet Radius of Curvature 1  	&  convex 61.10 mm	& convex 59.11 mm  \\ \hline
\rule[-1ex]{0pt}{2.5ex} Second Achromatic Doublet Radius of Curvature 2  	&  concave 18.76 mm	& concave 26.98 mm  \\ \hline
\rule[-1ex]{0pt}{2.5ex} Second Achromatic Doublet Radius of Curvature 3	&  concave 96.88 mm	& concave 256.96 mm  \\ \hline
\rule[-1ex]{0pt}{2.5ex} Second Achromatic Doublet First Lens Thickness	&  6.0 mm		&   \\ \hline
\rule[-1ex]{0pt}{2.5ex} Second Achromatic Doublet Second Lens Thickness	&  2.0 mm 		&   \\ \hline
\rule[-1ex]{0pt}{2.5ex} Second Achromatic Doublet First Lens Material 			&  \multicolumn{2}{|c|}{S-FPL51}   \\ \hline
\rule[-1ex]{0pt}{2.5ex} Second Achromatic Doublet Second Lens Material 		&  \multicolumn{2}{|c|}{S-NSL36}   \\ \hline
\rule[-1ex]{0pt}{2.5ex} Second Achromatic Doublet Diameter					&  16.0 mm 	& 16.21  \\ \hline
\rule[-1ex]{0pt}{2.5ex} Total Distance betw.~Second Achromat and Field Lens 	& 104.27 mm	& 126.51 mm  \\ \hline
\rule[-1ex]{0pt}{2.5ex} Distance betw.~Second Achromatic Doublet and Fold 	& varies 37.19--	& varies 64.87-- \vspace{-1pt} \\ 
\rule[-1ex]{0pt}{2.5ex} ~ ~Mirror \#1			&   	56.78 mm		& 78.81 mm \\ \hline
\rule[-1ex]{0pt}{2.5ex} Fold Mirror \#1 Physical Size			&  \multicolumn{2}{|c|}{23.0 mm $\times$~16.0 mm} \\ \hline
\rule[-1ex]{0pt}{2.5ex} Distance betw.~Fold Mirror \#1 and Fold Mirror \#2				&  varies 29.8--62.6 mm 	& varies 20.5--42.3 mm  \\ \hline
\rule[-1ex]{0pt}{2.5ex} Fold Mirror \#2 Physical Size: Circular Diameter			&   	18.70 mm 		& 18.90 mm  \\ \hline
\rule[-1ex]{0pt}{2.5ex} Fold Mirror \#2 Physical Size: Trimmed Height			&   	10.32 mm 		& 10.60 mm  \\ \hline
\rule[-1ex]{0pt}{2.5ex} Distance betw.~Fold Mirror \#2 and Field Lens			&   	varies 4.5--18.0 mm & varies 16.0--28.0mm  \\ \hline
\rule[-1ex]{0pt}{2.5ex} Field Lens Radius of Curvature 1			& convex 47.93 mm, 	& convex 57.22 mm, \vspace{-1pt} \\ 
\rule[-1ex]{0pt}{2.5ex} 											& 50.04 mm, and  		& 59.46 mm, and \vspace{-1pt} \\ 
\rule[-1ex]{0pt}{2.5ex} 											&  56.20 mm 			& 69.06 \\ \hline
\rule[-1ex]{0pt}{2.5ex} Field Lens Radius of Curvature 2			& flat  			& flat  \\ \hline
\rule[-1ex]{0pt}{2.5ex} Field Lens Thickness					& 2.00 mm  	& 2.00 mm  \\ \hline
\rule[-1ex]{0pt}{2.5ex} Field Lens Diameter						&  22.00 mm 	& 21.00 mm  \\ \hline
\rule[-1ex]{0pt}{2.5ex} Distance betw.~Field Lens and Focus	& 1.00 mm  	& 8.00 mm  \\ \hline
\end{tabular}
\end{center}
\end{table}

\bibliography{refs} 
\bibliographystyle{spiebib} 

\end{document}